\documentclass[a4paper, 12pt]{article}

\usepackage{epsfig,amssymb,euscript,xspace, color}
\usepackage{amsmath,jheppub,mathtools}
\def\bea{\begin{eqnarray}}
\def\eea{\end{eqnarray}}
\def\be{\begin{equation}}
\def\ee{\end{equation}}



\begin{document}

\title{BTZ dynamics  and chaos.  }
\author{Rohan R. Poojary$^{1,2}$}
\affiliation{$^1$Department of Theoretical Physics, \\
Tata Institute of Fundamental Research,\\
Mumbai 400005, India.\\ \\
$^2$Chennai Mathematical Institute,\\
H1-Sipcot IT Park, Siruseri 603103, India.
\\}
\emailAdd{rpglaznos@gmail.com,ronp@theory.tifr.res.in}
\date{}

\abstract{We find an effective action for gravitational interactions with scalars in $AdS_3$ to first order in $G_N$ at the conformal boundary. This action can be understood as an action for the Brown-Henneaux modes and is given by the square-root product of right and left moving Schwarzian derivatives for conformal transformations of the boundary. We thus reproduce the result  $\lambda_L=\frac{2\pi}{\beta}$ for OTOC computed first in  arXiv:1412.6087 for a Schwarzchild black hole in $AdS_3$. Applying the same procedure to rotating BTZ we find the Lyapunov index to be $\lambda_L=\frac{2\pi}{\beta_+}>\frac{2\pi}{\beta}$ where $\beta_+=\beta(1-\mu_L)$, with $\mu_L=\frac{r_-}{r_+}$ being the chemical potential for angular momentum. We thus comment on a possible generalization to a part of the proof given in arXiv:1503.01409 to accommodate this result.  
\\\\
}
\maketitle


\section{Introduction}
A very intriguing phenomena of strongly coupled thermal systems is chaos. In a classical sense, phase space trajectories which differ in their initial values by a small amount tend to grow exponentially far apart at later times.  A good quantum mechanical analogue of this would be the time scale when
\begin{equation}
C(t)=\langle[W(t),V(0)]^2\rangle_\beta
\label{otoc1}
\end{equation}  
becomes equal to $ 2 \langle WW\rangle_\beta\langle VV\rangle_\beta$. Here, $W \& V$ are simple Hermitian operators with $\mathcal{O}(1)$ degrees of freedom. The exponential increase in the time of $C(t)\approx e^{\lambda_L t}$ can be considered as the Lypunov index generically associated with chaotic systems \cite{Kitaev:2014talk}. The chaotic behaviour of thermal large $N$ CFTs is related to the chaotic behaviour of black holes $via$ the gauge-gravity duality, the latter   are conjectured to be the fastest "scramblers" of information \cite{Sekino:2008he}. Shenker and Stanford \cite{Shenker:2014cwa} first computed the $out$-$of$-$time$-$ordered$ (otoc) term in (\ref{otoc1})
\begin{equation}
\langle W(t)V(0)W(t)V(0) \rangle_\beta
\label{otoc2}
\end{equation}
holographically using the eikonal approximation. In \cite{Shenker:2014cwa} the next order in probe approximation in $G_N$ was computed for a $2\rightarrow2$ scattering for 2 minimally coupled scalars in $AdS_d$ Schwarzchild interacting with each other only $via$ gravity. The Lyapunov index thus obtained was $\lambda_L=2\pi/\beta$, $\beta$ being the inverse temperature of the $AdS_d$ Schwarzchild. This lead Maldacena, Shenker and Stanford \cite{Maldacena:2015waa} to propose a bound on the Lyapunov index of large $N$ thermal QFTs to be $\lambda_L \leq 2\pi/\beta$ by using generic arguments of unitarity and analyticity of Whightman functions on the complex plane. It was assumed that holographic CFTs saturate this bound as evidenced by \cite{Shenker:2014cwa,Shenker:2013pqa}.
\\\\
Further interest in chaotic systems was heightened by the study of the SYK \cite{Sachdev:2010um}and SYK-like models initiated first in \cite{Polchinski:2016xgd} and \cite{Maldacena:2016hyu}; in \cite{Maldacena:2016hyu} Maldacena and Stanford found that the otoc for the fermions has $\lambda_L=2\pi/\beta$, this computation was done in the strong coupling (zero temperature) limit  of the SYK model where the model is conformal. In order to compute the leading contribution to the 4pt. function they had to break the conformal invariance at zero temperature. The modes which are responsible for maximizing chaos where shown to be the modes related by diffeomorphism which now have an action due to breaking of conformal invariance. Their effective action was computed and found to be the Schwarzian derivative of reparametrizations of the thermal circle. Many interesting properties of the SYK model have since been uncovered \cite{Kitaev:2017awl,Sonner:2017hxc,Eberlein:2017wah,Gross:2017vhb,Dartois:2017xoe,Garcia-Garcia:2016mno,Gross:2017aos,Stanford:2017thb} to quote a few. 
\\\\
There have been many variations to the original SYK problem which had relied upon averaging over a space of couplings. A unitary model proposed by Gurav\cite{Gurau:2016lzk} and Witten\cite{Witten:2016iux} showed a similar behaviour to the SYK model at large $N$. There have also since been higher-dimensional and super-symmetric avatars of this model, \cite{Choudhury:2017tax,Gonzalez:2018enk,Bhattacharya:2017vaz,Krishnan:2017lra,Yoon:2017nig,Bulycheva:2017uqj,Murugan:2017eto,Li:2017hdt,Davison:2016ngz,Klebanov:2016xxf,Berkooz:2016cvq,Fu:2016vas,Gross:2016kjj,Gu:2016oyy} study interesting properties.
\\\\
This lead to investigations to ascertain the bulk degrees of freedom which are responsible for similar chaotic behaviour. The dynamics of near extremal black holes was found to be captured by a 2d dilaton-gravity theory of Jackiw\cite{Jackiw:1984je} and Teitelboim\cite{Teitelboim:1983ux} in \cite{Almheiri:2014cka}. The   $nAdS_2$ dynamics of  Jackiw-Teitelboim (JT) action is essentially dictated by its asymptotic symmetries since the theory possesses zero propagating degrees of freedom. The effective action for these modes was captured by the Schwarzian action for the $AdS_2$ boundary diffeomorphisms \citep{Jensen:2016pah,Maldacena:2016upp}, similar ideas where pursued in \cite{Engelsoy:2016xyb}. Explicit computations on near extremal RN $AdS_4$ black holes \cite{Nayak:2018qej} corroborated this understanding from a higher dimensional perspective.  
\\\\
The dynamics of the 2d gravity theory reproducing the Schwarzian effective action have since been studied 
\cite{Mertens:2018fds,Taylor:2017dly}. Apart from the JT action, the Polyakov action with a cosmological constant was also studied and found to describe the $AdS_2$ bulk dynamics dual to the soft modes of SYK \cite{Mandal:2017thl}. This was done by analysing the action for the co-adjoint orbits of the Virasoro group which can be thought to describe the soft modes. Different aspects of $AdS_2$ gravity were also covered in  
\cite{Haehl:2017pak,Grumiller:2017qao,Dubovsky:2017cnj,Eling:2017txo,Kyono:2017pxs,Forste:2017kwy,Cvetic:2016eiv,Almheiri:2016fws,Engelsoy:2016xyb,Gaikwad:2018dfc}. The effect of $AdS_2$ arising in rotating horizons was also studied in \cite{Anninos:2017cnw}, here a large $N$ SYK like system was modelled to mimic the near horizon near extremal symmetries of Kerr-Neuman blackholes in $AdS_4$. For past works the reader may refer to \cite{Gouteraux:2011qh,Castro:2008ms} and references therein. 
\\\\
There have also been efforts to realize the SYK model completely by a providing a holographic description \cite{Jevicki:2016bwu,Jevicki:2016ito,Das:2017pif,Das:2017hrt,Das:2017wae}. These have also been studied for the SYK tensor models too in some detail \cite{Forste:2017apw,Halmagyi:2017leq,Cai:2017nwk,Caputa:2017yrh,Krishnan:2017txw,Gross:2017hcz,Krishnan:2016bvg}. 
\\\\
It is also worth noting that a theory of open string governed by the Nambu-Goto action  probing an $AdS$ Schwarzchild geometry also exhibits maximal chaos \cite{deBoer:2017xdk}. In such a system the scrambling time is governed by the string tension. A Schwarzian effective action has also been uncovered for such systems as being comprised of the reparametrizations of the world sheet \cite{Banerjee:2018twd,Banerjee:2018kwy}.  
\\\\
It would be an interesting question to ask if such modes can be found in thermal large $N$ CFTs such that their effective actions govern the chaotic behaviour of the system, like in the SYK model studied in \cite{Maldacena:2016hyu}. The present holographic understanding of this phenomenon allows one to visualize these modes close to extremality in the near horizon region for at least non-rotating geometries.  The near extremal geometries possess a near horizon $AdS_2$ throat, and bulk scattering of the form studied in \cite{Shenker:2014cwa} excite graviton modes which can be described by a JT theory confined to this throat region \citep{Maldacena:2015waa}.  It is worth noting that the holographic computations in \citep{Shenker:2014cwa,Shenker:2013pqa} do not assume extremality. It would therefore be worthwhile to understand how these modes behave away from extremality and also in the entirety of a black hole in $AdS$. 
\\\\
To this end we address a simpler problem of that in $AdS_3$ which like in the dilaton-gravity theory in $AdS_2$ has only boundary degrees of freedom. In fact in $AdS_3$ these have been well studied and are called the Brown-Henneaux modes \cite{Brown:1986nw} which are in one-to-one correspondence with 2d infinite conformal symmetries of the boundary CFT$_2$. In section 2 using the gauge gravity prescription we first formally equate the computation of eikonal scattering in the bulk done in \cite{Shenker:2014cwa} to computing correlators in the boundary CFT upto linear order in $G_N$. This we do by computing the effective action for conformal transformations on the boundary obtained from the bulk on-shell path integral. We then (section 3) compute the effective action for the Brown-Henneaux modes about a rotating BTZ and find it to be the product of square-root Schwarzian derivatives, each for left and right moving conformal transformations of the boundary.
\\\\
In section 4 we proceed to compute the correction to the 4pt function of 2 boundary operators -  computed in the probe approximation; to linear order in $G_N$. We thus reproduce the answer of \cite{Shenker:2014cwa} for the non-rotating BTZ case of $\lambda_L=2\pi/\beta$. The similar procedure when used for the rotating BTZ yields $\lambda_L=2\pi/\beta_+$ where 
$\beta_\pm=\beta(1\mp\mu_L)$ with $\mu_L=r_-/r_+$ being the chemical potential associated with angular momentum. We thus find that for the rotating BTZ $\lambda_L=2\pi/\beta_+>2\pi/\beta$.
\\\\
We end with section 5 with some conclusions and discuss the possible implications of the result. In particular we point out a possible modification of a part of the proof given in \cite{Maldacena:2015waa} so as to allow for a modified bound in the presence of a chemical potential for angular momentum.  
\section{Bulk Computation}
In this section we heuristically equate the eikonal approximate calculation of \cite{Shenker:2014cwa} to the one generally done while introducing the AdS/CFT correspondence $i.e.$ equating the bulk on-shell (small $G_N$) path-integral to the generating function of boundary correlators:
\begin{equation}
\underset{\phi\rightarrow\phi_0}{\underset{g\rightarrow\eta}{\int}}\mathcal{D}[g]\mathcal{D}[\phi_i]\,\,e^{i(S_{grav}+S_{matter})}=Z_{CFT}[\phi_0]=\langle e^{i\int_{\partial}\phi_0\mathcal{O}}\rangle_{CFT}
\label{adscft}
\end{equation}
where $\phi_0$ is the boundary value of the scalar field in the bulk which sources a scalar operator $\mathcal{O}$ in the boundary CFT.
Like in \cite{Shenker:2014cwa}
we will consider 2 minimally coupled scalar fields in the bulk with masses $m_1\&\,m_2$\footnote{There is summation in $i$ and the space time integrals are suppressed for brevity.}, with no interaction terms between them
\begin{equation}
S_{matter}=-\int\sqrt{-g}\tfrac{1}{2}\left[(\partial\phi_i)^2-m^2_i\phi_i^2\right],\,\hspace{1cm}\,S_{grav}=-\frac{1}{16\pi G_N}\int\sqrt{-g}(R-2\Lambda).
\end{equation}
Here we have not written down the boundary terms for the actions which make the variational problem well defined and render the on-shell action finite.
\\\\
We will concern ourselves with the computation of the bulk 4pt. function $\langle\phi_1\phi_1\phi_2\phi_2\rangle$, which is equal to the boundary 4pt. function
\begin{equation}
\langle\mathcal{O}_1\mathcal{O}_1\mathcal{O}_2\mathcal{O}_2\rangle\approx \lim_{r\rightarrow\infty}r^{-2(2d-\Delta_1-\Delta_2)}\langle\phi_1\phi_1\phi_2\phi_2\rangle
\label{bulkbndy1}
\end{equation}
where each of the bulk coordinates is taken to the boundary\footnote{ $\Delta_i=\tfrac{d}{2}+\sqrt{\tfrac{d^2}{4}+m_i^2l^2}$, where $l$ is the $AdS$ radius.}. Using the bulk path integral expression, the 4pt function is given by
\begin{equation}
\langle\phi_1\phi_1\phi_2\phi_2\rangle=\int\mathcal{D}[g]\mathcal{D}[\phi_i]\,\,\phi_1\phi_1\phi_2\phi_2 \,\,e^{i(S_{grav}+S_{matter})}
\label{bulkpi}
\end{equation}
Here for simplicity of notation we have not mentioned the space time dependence. Since we would be concerned with the limit in which classical gravity dominates we would be interested in the saddle point evaluation of the above path-integral. Further  one usually considers the probe approximation in which the scalars act as probes for a given metric which satisfies vacuum Einstein's equation with cosmological constant $\Lambda$. We will denote this solution as $\bar{g}_{\mu\nu}$. In this limit since the matter action is quadratic the answer is readily computed from the expression
\begin{equation}
\langle\phi_1\phi_1\rangle_{\bar{g}}\langle\phi_2\phi_2\rangle_{\bar{g}} =\int\mathcal{D}[\phi_i]\,\,\phi_1\phi_1\phi_2\phi_2 \,\,e^{i(S_{grav}[\bar{g}]+S_{matter}[\bar{\phi_i}])},
\label{bulkpi0}
\end{equation}
here $\bar{\phi_i}$ solves the Klein-Gordon equation in the background $\bar{g}_{\mu\nu}$\footnote{This is basically the expression for the propagator in the form of a path integral for a free theory. }. Let us consider the effect of first order back reaction in orders of $G_N$ by considering
\begin{equation}
R_{\mu\nu}-\frac{1}{2}Rg_{\mu\nu}+\Lambda g_{\mu\nu}=4\pi G_N T_{\mu\nu}
\label{bulkeom}
\end{equation}
where $g_{\mu\nu}=\bar{g}_{\mu\nu}+h_{\mu\nu}$ and $T_{\mu\nu}$ is determined entirely in terms of $\bar{\phi_i}$s. Therefore we can rewrite (\ref{bulkpi}) as
\begin{eqnarray}
\langle\phi_1\phi_1\phi_2\phi_2\rangle&=&\int\mathcal{D}[h]\mathcal{D}[\phi_i]\,\,\phi_1\phi_1\phi_2\phi_2\,\, e^{i\left\lbrace S_{grav}[\bar{g}]+\delta S_{grav}[h]+S_{matter}[\bar{g},\bar{\phi_i}]+\delta S_{matter}[h]\right\rbrace}\cr
&=&\int\mathcal{D}[h]\langle\phi_1\phi_1\rangle_{\bar{g}+h}\langle\phi_2\phi_2\rangle_{\bar{g}+h}\,\, e^{i\delta S_{grav}[h]}\cr
&=&\int\mathcal{D}[h]\langle\phi_1\phi_1\rangle_{\bar{g}+h}\langle\phi_2\phi_2\rangle_{\bar{g}+h}\,\, {\rm exp}\left[\tfrac{il}{16\pi G_N}\int \tfrac{1}{2}h{\rm D^2}h\right]
\label{bulkpi2}
\end{eqnarray}
where $\langle\phi_1\phi_1\rangle_{\bar{g}+h}$ denotes the 2pt. function evaluated in background metric $\bar{g}_{\mu\nu}+h_{\mu\nu}$ for an $h_{\mu\nu}$ determined by (\ref{bulkeom}) and constrained by boundary conditions on the bulk metric. Note, that here we have only considered diagrams in which the graviton lines attach to different scalar legs, corrections to scalar propagator due to gravitons attaching to the same scalar propagator have been ignored. 
\\\\
In the eikonal approximation $i.e.$ in the limit when the scalar field momenta are taken to be large and light-like, (\ref{bulkpi2}) reduces to \cite{Kabat:1992tb}
\begin{equation}
\int\mathcal{D}[h]\,\,{\rm exp}\left[\frac{il}{16\pi G_N}\int \tfrac{1}{2}hD^2h +h_{\mu\nu}T^{\mu\nu}\right]
\end{equation}
where we have assumed that the $T_{\mu\nu}$ due to the matter fields are like shockwaves with light-like momenta. In this approximation the incoming and the out-going momenta are the same and we get the effect of infinite graviton ladder exchanges between the 2 scalar propagators. This precisely the origin of the $e^{i\delta(s)}$ factor in \cite{Shenker:2013pqa}. Here $h_{\mu\nu}$ is the response to the shockwave generated by high momentum scalar propagators in $T_{\mu\nu}$.
This approximation is justified in the shock wave analysis of \cite{Shenker:2013pqa} since the scattering is arranged to have a maximum contribution from the near horizon region. By the time any scalar perturbation reaches the horizon area it would be blue-shifted exponentially, thus giving the first leading contribution to the correction to the 4pt function. The eikonal approximation used in \cite{Shenker:2014cwa} effectively means that the contribution comes from the bifurcate horizon. One can then use bulk to boundary propagators to  compute the correlation function on the boundary. 
\\\\
Let us now try to rephrase the same computation in $AdS_3$. We  write the probe approximate 4pt function as
\begin{equation}
\langle\mathcal{O}_1\mathcal{O}_1\rangle_{\bar{g}}\langle\mathcal{O}_2 \mathcal{O}_2\rangle_{\bar{g}}=\lim_{r\rightarrow\partial}\int\mathcal{D}[\phi_i]\,\,\phi_1\phi_1\phi_2\phi_2\,\, e^{iS_{matter}[\bar{g},\bar{\phi_i}]}
\label{4pt1}
\end{equation}
here we have haven't included $S_{grav}[\bar{g}]$ since it would a constant. The correction it would receive from the gravity path integral would be captured in 
\begin{equation}
\langle\mathcal{O}_1\mathcal{O}_1\mathcal{O}_2\mathcal{O}_2\rangle=\int \mathcal{D}[g] \langle\mathcal{O}_1\mathcal{O}_1\rangle_{g}\langle\mathcal{O}_2 \mathcal{O}_2\rangle_{g}\,\,{\rm exp}\left[\tfrac{il}{16\pi G_N}\int\sqrt{-g}(R-2\Lambda)\right].
\label{4pt2}
\end{equation}
We could now consider the above path integral being dominated by the gravity saddle by considering the small $G_N$ limit. Therefore we would only be seeking contributions from geometries satisfying vacuum Einstein's equations. This way we should be able to reproduce the leading correction in the small $G_N$ limit. 
In the above metric path integral only boundary degrees of freedom contribute in $AdS_3$. These are in one-to one correspondence with the boundary conformal transformations\footnote{Barring the truly small diffeomorphisms which we ignore.}. Further $\langle\mathcal{O}_1\mathcal{O}_1\rangle_{g}\approx\langle\mathcal{O}_1\mathcal{O}_1\rangle_{\bar{g}+h}$ would simply correspond to the change in the 2pt. functions due to conformal transformations. Therefore atleast in $AdS_3$ we must be able to capture the effect of chaos $via$ (\ref{4pt2}).

\section{$AdS_3$ story from the boundary}
In this section we will derive the effective action for the soft modes by calculating the gravity bulk on-shell action about an arbitrary BTZ geometry.
As justified in the previous section this amounts to finding the effective action for the diffeomorphisms in the bulk respecting Dirichlet boundary conditions. It is known that such configurations in the bulk are dual to states in the boundary CFT which are created by the action of 2 copies of commuting Virasoro algebra. We begin by writing the most general bulk configuration with the boundary metric being flat, we begin with a Lorenztian metric in the Fefferman-Graham gauge \cite{Banados:1998gg,Henningson:1998ey}.
\bea
\frac{ds^2}{l^2}&=&\frac{dr^2}{r^2}-\frac{r^2dx^+dx^-}{4}+\frac{1}{4}\left(T_{++}dx^{+2}+T_{--}dx^{-2}\right)-\frac{1}{4r^2}T_{++}T_{--}dx^+dx^-,
\label{Banados_metric}
\eea 
where $T_{++}=T_{++}(x^+)\,\,T_{--}=T_{--}(x^-)$ and $x^\pm=t\pm\phi$. One can cast the BTZ \cite{Banados:1992wn} metric in $AdS_3$
\bea
\frac{ds^2}{l^2}&=&\frac{r^2dr^2}{(r^2-r_+^2)(r^2-r_-^2)}-\frac{(r^2-r_+^2)(r^2-r_-^2)dt^2}{r^2}+r^2\left(d\phi-\frac{r_+r_-}{r^2}dt\right)^2\cr&&\cr
M&=&r_+^2+r_-^2,\,\,J=2lr_+r_-.
\label{BTZ_metric}
\eea
in the Fefferman-Graham gauge with $T_{\pm\pm}=(r_+\pm r_-)^2$. It is worthwhile to notice that the radial coordinate in \eqref{Banados_metric} sees the horizon at $r_h=\sqrt{r_+^2-r_-^2}\,$\footnote{Throughout the text $r_\pm$ would refer to outer or inner horizons in the metric in \eqref{BTZ_metric}and will be casual in the use of $r$ as the radial coordinate in any metric.  }. \\\\
The bulk action with relevant boundary(counter) terms is\cite{Henningson:1998ey,deHaro:2000vlm}
\bea
16\pi G_N S_{bulk}&=&\int d^3x \sqrt{-g}(R-2\Lambda)+2\int_{\partial}d^2x\sqrt{-h}\left(K+\frac{1}{l}\right),
\label{bulk_action}
\eea 
where $\Lambda=-1/l^2$ and $l$ is the length of $AdS_3$. The $1/l$ term in the boundary action is used to make the on-shell action finite. Therefore the on-shell value of \eqref{bulk_action} for arbitrary metrics given by \eqref{Banados_metric} is
\bea
S^{on-shell}_{bulk}&=&\frac{l}{64\pi G_N}\int_\partial\left(r_h^2+\frac{T_{++}T_{--}}{r_h^2}\right)\cr
&=&\frac{l}{32\pi G_N}\int_\partial\sqrt{T_{++}T_{--}}
\label{bulk_onshell_action}
\eea 
Since gravity in 3-dim is non-dynamical in the bulk, all solutions to the bulk action \eqref{bulk_action} can be obtained from one another $via$ diffeomorphisms. In fact the Fefferman-Graham theorem \cite{Henningson:1998ey} allows one to express any solution in the form \eqref{Banados_metric} as a diffeomorphisms of the other and we thus only need to concern ourselves with diffeomorphisms that preserve this form to generate all solutions. We would therefore like to know the action \eqref{bulk_onshell_action} associated with such a diffeomorphism, having started from a particular solution (for eg.) of the form of the BTZ metric \eqref{BTZ_metric}. To this end we would like to know how $T_{\pm\pm}$ depend on these diffeomorphisms. 
\\\\
We notice that for infinitesimal diffeomorphisms which maintain the form of \eqref{Banados_metric}, the change in $T_{\pm\pm}$ is given by \cite{Balasubramanian:1999re,Brown:1986nw}
\bea
\delta T_{\pm\pm}&=&\xi^\pm_{(0)}T'_{\pm\pm}+2{\xi_{{0}}^\pm}'T_{\pm\pm}-2{\xi_{{0}}^\pm}'''
\label{delta_T}
\eea
where\footnote{The primes denote derviatives $w.r.t.$ respective coordinate dependence.}
\bea
\xi^\mu\partial_\mu&=&\xi^r\partial_r+\xi_{(0)}^+(x^+)\partial_+ + \xi_{(0)}^-(x^-)\partial_-+{\mathcal{O}(1/r)}\cr
\xi^r&=&-\frac{r}{2}\left({\xi_{(0)}^+}'+{\xi_{(0)}^-}'\right).
\label{FG_diffeomorphism}
\eea    
We also note that change in a Schwarzian derivative $\{T(u),u\}$ due to a diffeo $u\rightarrow u+ \epsilon(u)$ is:
\bea
\{T(u)+\epsilon(u)T'(u),u\}&=&\{T,u\}+\epsilon(u)\partial_u\{T,u\}+2\epsilon'(u)\{T,u\}+\epsilon'''(u).
\label{delta_Sch}
\eea
One can find the full non-linear completion of (\ref{FG_diffeomorphism}) \cite{Roberts:2012aq} which takes a Poinc\'{a}re $AdS_3$ (with $T_{\pm\pm}=0$) to (\ref{Banados_metric}). Under such a diffeomorphism the stress tensor is proportional to the Schwarzian for boundary conformal transformations. Comparing \eqref{delta_T} and \eqref{delta_Sch} we can deduce that under $x^\pm\rightarrow X^\pm(x^\pm)$
\bea
T_{\pm\pm}&=&-2\{X^\pm,x^\pm\}\cr
{\rm where \,\,\,}\{X,x\}&=&\frac{2X'X'''-3X''^2}{2X'^2}
\eea
where for infinitesimal diffeomorphisms $X^\pm\equiv x^\pm + \xi_{(0)}^\pm$. One notices that for $X^\pm=x^\pm\implies$ $T_{\pm\pm}=0$. This value can be shifted by defining
\bea
T_{\pm\pm}&=&-2\{X^\pm,x^\pm\}+L_\pm{X^\pm}'^2
\eea
where the choice of ${X^\pm}'^2$ makes sure that the linear in $T_{\pm\pm}$ terms in \eqref{delta_T} remain the same. Here, $L_\pm$ define the charge of the BTZ metric about which the change in the parameters $T_{\pm\pm}$ is measured\footnote{$T_{\pm\pm}$ are components of the Brown-York stress tensor for the bulk metric, which are also the CFT$_2$ stress tensor components.}.
\\\\
Therefore the on-shell action in \eqref{bulk_onshell_action} is
\bea
S^{on-shell}_{bulk}&=&\frac{l}{32\pi G_N}\int_\partial\sqrt{\left(-2\{X^+,x^+\}+L_+{X^+}'^2\right)\left(-2\{X^-,x^-\}+L_-{X^-}'^2\right)}
\label{bulk_onshell_action_explicit}
\eea
where $L_\pm$ decides which bulk configuration one mesures the change from. The above action is defined on the boundary of $AdS_3$, the integral in the $AdS$ radial direction receives contribution from $r=r_h$ and the boundary $r=\infty$. The divergent contribution from the boundary at $r=\infty$ is cancelled by the holographic boundary counter-terms, the only finite contribution comes from the horizon. Under infinitesimal diffeomorphisms $X^\pm\rightarrow x^\pm + \epsilon^\pm(x^\pm)$ the quadratic action takes the form
\bea
&&S^{on-shell}_{bulk}=\tfrac{-l}{64\pi G_N(L_+L_-)^{3/2}}\int_\partial\left(L_-^2({\epsilon^+}'''(x^+)^2+L_+{\epsilon^+}''(x^+)^2)+L_+^2({\epsilon^-}'''(x^-)^2+L_-{\epsilon^-}''(x^-)^2)\right)\cr&&
\label{bulk_onshell_action_quad}
\eea
where we have ignored boundary terms. Since we would be interested in computing OTOC's later we Euclideanize the above action
\bea
&&S^{on-shell}_{bulk,E}=\tfrac{l}{64\pi G_N(L_+L_-)^{3/2}}\int_\partial\left(L_-^2({\epsilon^+}'''(x^+)^2-L_+{\epsilon^+}''(x^+)^2)+L_+^2({\epsilon^-}'''(x^-)^2-L_-{\epsilon^-}''(x^-)^2)\right)\cr&&
\label{bulk_onshell_action_quad_euclid}
\eea
The quadratic action divides itself into left and right sector. 
The action (\ref{bulk_onshell_action_explicit}) evidently has the symmetries of the Schwarzian derivative, the infinitesimal versions of which are manifested in (\ref{bulk_onshell_action_quad_euclid}). We would correspondingly have got the above action by working in the bulk in a Euclidean setting to begin with. This would have invariably required us to have the angular momentum associated with the BTZ metric to be imaginary so as to have a real Euclidean metric.
\\\\
One can in principle derive the one-loop exact action for the boundary gravitons in $AdS_3$ as was done recently by \cite{Cotler:2018zff} using the Chern-Simmons prescription. The authors obtain a theory of re-parametrizations which encodes loop contributions of boundary gravitons in a perturbation in $1/c$\footnote{$c$ being the central charge.}. 
\subsection{Propagators}
We are now in the Euclidean setting where  the time $\tau=-it$ is along the imaginary direction while the space-like coordinate $\phi$ is real. The $x^\pm$ coordinates of Euclideanised BTZ metric can be regarded to have complex periodicities \cite{KeskiVakkuri:1998nw}
\be
x^\pm = x^\pm+i\beta_\pm,\,\,\,\,\,\beta_\pm=\beta\mp i\Omega=\frac{2\pi}{\sqrt{L_\pm}}
\label{periodicity}
\ee 
We regard the integral in (\ref{bulk_onshell_action_quad_euclid}) to be in one such periodic interval, therefore the above action splits into two 1-dim actions
\be
S_+[\epsilon^+]=\alpha_+\int \epsilon^+(\bar{\partial}^{(6)}+L_+\bar{\partial}^{(4)})\epsilon^+,\,\,\,
S_-[\epsilon^-]=\alpha_-\int \epsilon^-(\partial^{(6)}+L_-\partial^{(4)})\epsilon^- ,
\ee  
where $\alpha_\pm = \frac{-2\pi l}{64\pi G_N L_\pm^{3/2}}$. For convenience we have defined
\be
\bar{z}=-ix^+=\bar{z}+\beta_+,\,\,\,z=-ix^-=z+\beta_-
\label{periodicity1}
\ee
and we analyse the propagator for $\epsilon^+$, for which we evaluate the Green's function $G_+$ for the operator
\be
\mathbb{O}_+=\bar{\partial}^{(4)}\left(\bar{\partial}^{(2)}+\left(\tfrac{2\pi}{\beta_+}\right)^2\right),\,\,\,\mathbb{O}_+G_+=\delta(\bar{z}).
\label{propagatoreq}
\ee
Here, we observe that $G_+$ would depend on $\bar{z}$ by a function of the ratio $\bar{z}/\beta_+$. The zero modes themselves would look like $\{1,e^{\tfrac{2\pi\bar{z}}{\beta_+}},e^{-\tfrac{2\pi\bar{z}}{\beta_+}}\}$. We will compute the  $G_+$ first for the Schwarzchild case and try and generalize for the rotating BTZ case.
\\\\
For $AdS_3$ Schwarzchild, $\beta_+=\beta_-=\beta\in{\mathbb{R}}$. Assuming $G_+$ be a function of $\bar{z}/\beta$ we solve (\ref{propagatoreq}) for real values of $\bar{z}/\beta$ $i.e.$ $G_+(\tau/\beta)$ and then using Schwarz's theorem analytically continue it for arbitrary complex values of $\bar{z}/\beta$ \cite{Streater:1964}.
This allows us to express both sides of the (\ref{propagatoreq}) as a discrete sum, thus
\be
G_+=\frac{1}{\alpha_+}{\sideset{}{'}\sum_{n=-\infty}^{\infty}}\frac{e^{2\pi i n \bar{z}/\beta}}{\left(\tfrac{2\pi}{\beta}\right)^6 n^4 (1-n^2)},\hspace{0.5cm}\forall\,\, \frac{\bar{z}}{\beta}\in {\mathbb{R}},
\ee
where the prime on the sum denotes $n\notin\{-1,0,1\}$. Doing the relevant Matsubara summation and analytically continuing in the complex $\bar{z}/\beta$ plane yields
\bea
\left(\tfrac{\pi^3 l}{4G_N\beta^3}\right)G_+(\bar{z})&=& \tfrac{1}{24}\left(2\pi \|\tfrac{\bar{z}}{\beta}\|-\pi \right)^4 -\tfrac{(\pi^2+6)}{12}\left(2\pi \|\tfrac{\bar{z}}{\beta}\|-\pi \right)^2+\cr
&&\hspace{0.1cm}+\pi\left(2\pi \|\tfrac{\bar{z}}{\beta}\|-\pi \right)\sin\left(2\pi \|\tfrac{\bar{z}}{\beta}\|\right)+
+a+b\cos\left(2\pi \|\tfrac{\bar{z}}{\beta}\|\right).
\label{propagator}
\eea
 Here,
\[
\Big\|\frac{\bar{z}}{\beta}\Big\|=
\begin{cases}
\frac{\bar{z}}{\beta}, & \text{if Re} \left[\frac{\bar{z}}{\beta}\right]>0\\&\\
\frac{-\bar{z}}{\beta}, & \text{if Re} \left[\frac{\bar{z}}{\beta}\right]<0.
\end{cases}
\]
The last 2 terms\footnote{Explicitly: $a=\left(1+\tfrac{\pi^2}{6}+\tfrac{7\pi^4}{360}\right)$ \& $b=9/2$. } in (\ref{propagator}) are comprised of the zero modes we neglected in the sum and would drop out of any computation which respects the bulk isometries. An identical expression would exist for $G_-$ in terms of $z/\beta$.
\subsection{Rotating BTZ propagators}
One could extend the above method naively to rotating BTZ case. This would require solving the (\ref{propagatoreq}) for the real values of $\bar{z}/\beta_+$
\be
\bar{z}/\beta_+\in \mathbb{R}\implies
\frac{(\tau+\mu \phi)+i(\mu \tau-\phi)}{\beta(1+\mu^2)}
\in\mathbb{R}\implies\phi=\mu\tau,
\label{realsection}
\ee
where $\Omega=\mu\beta$. Same holds true for $z/\beta_-\in{\mathbb{R}}$. It is quite clear from the outset that one could define Euclidean coordinates $\{\tilde{\tau},\tilde{\phi}\}$.
\begin{equation}
\tilde{\tau}=\frac{\tau+\mu\phi}{(1+\mu^2)},\hspace{1cm}\tilde{\phi}=\frac{\phi-\mu\tau}{(1+\mu^2)}.
\label{tildecoord}
\end{equation}
Therefore $\bar{z}/\beta_+=\bar{\tilde{z}}/\beta$, where $\bar{\tilde{z}}=\tilde{\tau}-i\tilde{\phi}$. Which is a conformal transformation on the boundary metric:
\begin{equation}
ds^2=d\tau^2+d\phi^2\rightarrow(1+\mu^2)(d\tilde{\tau}^2+d\tilde{\phi}^2).
\end{equation}
We do not however perform such a transformation on the propagator, we merely use (\ref{tildecoord}) for making the coordinated dependence look simple. Therefore solving (\ref{propagatoreq}) for real values of $\bar{\tilde{z}}/\beta$ and analytically continuing we get
\bea
\left(\tfrac{\pi^3 l}{4G_N\beta^3_+}\right)G_+(\bar{z})&=& \tfrac{1}{24}\left(2\pi \|\tfrac{\bar{z}}{\beta_+}\|-\pi \right)^4 -\tfrac{(\pi^2+6)}{12}\left(2\pi \|\tfrac{\bar{z}}{\beta_+}\|-\pi \right)^2+\cr
&&\hspace{0.1cm}+\pi\left(2\pi \|\tfrac{\bar{z}}{\beta_+}\|-\pi \right)\sin\left(2\pi \|\tfrac{\bar{z}}{\beta_+}\|\right)+
+a+b\cos\left(2\pi \|\tfrac{\bar{z}}{\beta_+}\|\right).
\label{propagatorBTZ}
\eea
where
\[
\Big\|\frac{\bar{z}}{\beta_+}\Big\|=
\begin{cases}
\frac{\bar{z}}{\beta_+}, & \text{if Re} \left[\frac{\bar{z}}{\beta_+}\right]>0\\&\\
\frac{-\bar{z}}{\beta_+}, & \text{if Re} \left[\frac{\bar{z}}{\beta_+}\right]<0.
\end{cases}
\]
Note that the conformal transformation (\ref{tildecoord}) isn't one of the $SL(2,\mathbb{R})$ zero modes.
\section{4pt correlator}
In this section we will use the propagators obtained in the last section to compute the next to leading order in $G_N$ corrections to the 4pt. function. We consider the first the leading contribution to the Euclidean 4pt. function of four scalars \cite{KeskiVakkuri:1998nw}
\be
\langle V_1V_2W_3W_4\rangle = \frac{1}{\sin^{2\bar{h}
_1}\left(\frac{\pi \bar{z}_{12}}{\beta_+}\right)\sin^{2h_1}\left(\frac{\pi z_{12}}{\beta_-}\right)\sin^{2\bar{h}2}\left(\frac{\pi \bar{z}_{34}}{\beta_+}\right)\sin^{2h_2}\left(\frac{\pi z_{34}}{\beta_-}\right)}
\label{4pt0}
\ee
where $z_{12}=z_1-z_2$ and $V_1\equiv V(\bar{z}_1,z_1)$. We would be interested in seeing how they would depend on the bulk on-shell metrics in the path integral (\ref{4pt2}). As explained before, these would correspond to computing the change in (\ref{4pt0}) due to conformal transformations parametrized by $\epsilon^+(\bar{z})$ \& $\epsilon^-(z)$. Under $\bar{z}\rightarrow\bar{z}+\epsilon^+(\bar{z})$ \& $z\rightarrow z+\epsilon^-(z)$ we have
\bea
\frac{1}{\sin^{2\bar{h}}\left(\frac{\pi \bar{z}_{12}}{\beta_+}\right)\sin^{2h}\left(\frac{\pi z_{12}}{\beta_-}\right)}&\rightarrow& \mathcal{B}(\epsilon^\pm_1,\epsilon^\pm_2)\frac{1}{\sin^{2h}\left(\frac{\pi \bar{z}_{12}}{\beta_+}\right)\sin^{2\bar{h}}\left(\frac{\pi z_{12}}{\beta_-}\right)},\cr
\mathcal{B}(\epsilon^\pm_1,\epsilon^\pm_2)&=& \bar{h}\left[(\epsilon^{+'}_1+\epsilon^{+'}_2)-\left(\frac{2\pi}{\beta_+}\right)\frac{(\epsilon^+_1-\epsilon^+_2)}{\tan\left(\frac{\pi \bar{z}_{12}}{\beta_+}\right)}\right]+{\rm c.c}
\label{2ptchange}
\eea
It can be seen that $\mathcal{B}$ above is invariant under the $SL(2,{\mathbb{R}})$ zero modes of  $\epsilon^+=\{1,e^{\pm 2\pi i \bar{z}/\beta_+} \}$ \& $\epsilon^-=\{1,e^{\pm 2\pi i z/\beta_-} \}$.
The correction to (\ref{4pt0}) is obtained by Wick contracting the $\epsilon^\pm$s with each other using the propagator (\ref{propagator}) and it's complex conjugate. We will first analyse this around $AdS_3$ Schwarzchild and then in a generic rotating BTZ background.

\subsection{$AdS_3$ Schwarzchild}\label{non_rotating_btz}
For the case of $AdS_3$ Schwarzchild we have $\beta_\pm=\beta$. The reality condition of (\ref{realsection}) implies $\phi=0$, $i.e.$ we compute the propagators $G_\pm$ along the $\tau$ real line and then analytically continue. Here to proceed we first have to order the Euclidean times for the operators in question and then use the appropriate propagator value for Wick contraction. We then add arbitrary Lorenztian time arguments corresponding to each operator and then read off the answer. The Euclidean answer to the expression
\be
\frac{\langle V_1 V_2 W_3 W_4 \rangle_{grav}}{\langle V_1 V_2 \rangle\langle W_3 W_4\rangle}=
\langle\mathcal{B}(\epsilon^\pm_1,\epsilon^\pm_2)\mathcal{B}(\epsilon^\pm_3,\epsilon^\pm_4)\rangle
\label{BB}
\ee 
looks like
\bea
h_1h_2&&\left\lbrace\tfrac{8}{3}\left(\tfrac{\pi}{\beta}\right)^6(z_{13}^4+z_{24}^4-z_{14}^4-z_{23}^4)\cot\left(\tfrac{\pi z_{12}}{\beta}\right)\cot\left(\tfrac{\pi z_{34}}{\beta}\right)\right.\cr
&&\hspace{0.2cm}+\tfrac{8}{3}\left(\tfrac{\pi}{\beta}\right)^5\left[ -2\pi (z_{13}^3+z_{24}^3-z_{14}^3-z_{23}^3)\cot\left(\tfrac{\pi z_{12}}{\beta}\right)\cot\left(\tfrac{\pi z_{34}}{\beta}\right)\right.\cr
&&\hspace{2cm}\left.+2(z_{24}^3+z_{14}^3-z_{23}^3-z_{13}^3)\cot\left(\tfrac{\pi z_{34}}{\beta}\right)-2(z_{24}^3-z_{14}^3+z_{23}^3-z_{13}^3)\cot\left(\tfrac{\pi z_{12}}{\beta}\right)\right]\cr
&&\hspace{0.2cm}+\tfrac{4}{3}\left(\tfrac{\pi}{\beta}\right)^4\left[-6(z_{13}^2+z_{14}^2+z_{23}^2+z_{24}^2)+(12-4\pi^2)z_{12}z_{34}\cot\left(\tfrac{\pi z_{12}}{\beta}\right)\cot\left(\tfrac{\pi z_{34}}{\beta}\right)\right.\cr
&&\hspace{2.4cm}\left.-12\pi(z_{13}+z_{24})\left(z_{12}\cot\left(\tfrac{\pi z_{12}}{\beta}\right)+z_{34}\cot\left(\tfrac{\pi z_{34}}{\beta}\right)\right)\right]\cr
&&\hspace{0.2cm}+\tfrac{8}{3}\left(\tfrac{\pi}{\beta}\right)^3\left[6\pi(z_{13}+z_{24})+2(\pi^2-3)\left(z_{12}\cot\left(\tfrac{\pi z_{12}}{\beta}\right)+z_{34}\cot\left(\tfrac{\pi z_{34}}{\beta}\right)\right)\right]\cr
&&\hspace{0.2cm}\left.+\tfrac{16}{3}\left(\tfrac{\pi}{\beta}\right)^2(\pi^2-3)\right\rbrace + {\rm c.c.}\Big|_{h_1\rightarrow\bar{h}_1,h_2\rightarrow\bar{h}_2}
\label{EuclideanTOSch}
\eea
Introducing Lorentzian time coordinates for each of operator $i.e.$ $z\rightarrow \tau +i \phi -it$, we fix $\tau_1=\beta,\tau_2=\beta/4,\tau_3=\beta/2,\tau_4=3\beta/4$ and further fix $t_1=t_2=t, t_3=t_4=0$ \& $\phi_1=\phi_2=0,\phi_3=\phi_4=\phi$. Having done this (\ref{EuclideanTOSch}) shows a polynomial growth in time.
\\\\
For the case of OTOC, contracting $\epsilon_i$ in (\ref{BB}) we similarly get
\bea
&&\tfrac{8}{3}\left(\tfrac{\pi}{\beta}\right)^6(z_{13}^4+z_{24}^4-z_{14}^4-z_{23}^4)\cot\left(\tfrac{\pi z_{12}}{\beta}\right)\cot\left(\tfrac{\pi z_{34}}{\beta}\right)\cr
&&\hspace{0.2cm}+\tfrac{8}{3}\left(\tfrac{\pi}{\beta}\right)^5\left[ -2\pi (z_{13}^3+z_{24}^3-z_{14}^3+z_{23}^3)\cot\left(\tfrac{\pi z_{12}}{\beta}\right)\cot\left(\frac{\pi z_{34}}{\beta}\right)\right.\cr
&&\hspace{2cm}\left.+2(z_{24}^3+z_{14}^3-z_{23}^3-z_{13}^3)\cot\left(\tfrac{\pi z_{34}}{\beta}\right)-2(z_{24}^3-z_{14}^3+z_{23}^3-z_{13}^3)\cot\left(\tfrac{\pi z_{12}}{\beta}\right)\right]\cr
&&\hspace{0.2cm}+\tfrac{4}{3}\left(\tfrac{\pi}{\beta}\right)^4\left[-6(z_{13}^2+z_{14}^2+z_{23}^2+z_{24}^2)+(12-4\pi^2)z_{12}z_{34}\cot\left(\tfrac{\pi z_{12}}{\beta}\right)\cot\left(\tfrac{\pi z_{34}}{\beta}\right)\right.\cr
&&\hspace{2.4cm}\left.-12\pi(z_{13}^2+z_{12}z_{34})\cot\left(\tfrac{\pi z_{12}}{\beta}\right)-12\pi (z_{24}^2+z_{12}z_{34})\cot\left(\tfrac{\pi z_{34}}{\beta}\right)\right]\cr
&&\hspace{0.2cm}+\tfrac{8}{3}\left(\tfrac{\pi}{\beta}\right)^3\left[6\pi\left(z_{14}+z_{23}\cot\left(\tfrac{\pi z_{12}}{\beta}\right)\cot\left(\tfrac{\pi z_{34}}{\beta}\right)\right)+2(\pi^2-3)\left(z_{12}\cot\left(\tfrac{\pi z_{12}}{\beta}\right)+z_{34}\cot\left(\tfrac{\pi z_{34}}{\beta}\right)\right)\right]\cr,
&&\hspace{0.2cm}+\tfrac{8}{3}\left(\tfrac{\pi}{\beta}\right)^2\left[-2(\pi^2-3)+3\pi\left(\tfrac{\sin\left(\tfrac{\pi(z_{12}+z_{34})}{\beta}\right)+\sin\left(\tfrac{\pi(z_{13}+z_{24})}{\beta}\right)}{\sin\left(\tfrac{\pi z_{12}}{\beta}\right)\sin\left(\tfrac{\pi z_{34}}{\beta}\right)}\right)\right] + \text{c.c}\Big|_{h_1\rightarrow\bar{h}_1,h_2\rightarrow\bar{h}_2}
\label{EuclideanOTOSch}
\eea
Similarly we introduce Lorentzian time coordinates for each of the operators $i.e.$ $z\rightarrow \tau +i \phi -it$, we fix $\tau_1=\beta,\tau_3=\beta/4,\tau_2=\beta/2,\tau_4=3\beta/4$ and further fix $t_1=t_2=t, t_3=t_4=0$ \& $\phi_1=\phi_2=0,\phi_3=\phi_4=\phi$. Here one clearly sees the exponential behaviour of the correlator for both the boundary null coordinates $t\pm\phi$ as
\be
\frac{\langle V_1(t,0) W_3(0,\phi) V_2(t,0) W_4(0,\phi) \rangle_{grav}}{\langle V_1(t,0) V_2(t,0) \rangle\langle W_3(0,\phi) W_4(0,\phi)\rangle}\sim \frac{G_N\beta}{ l}\left[h_1h_2 \cosh\left(\tfrac{2\pi(t-\phi)}{\beta}\right)+\bar{h}_1\bar{h}_2\cosh\left(\tfrac{2\pi(t+\phi)}{\beta}\right)\right]
\label{SchcldChaos}
\ee
Thus we see that the Schwarzian action (\ref{bulk_onshell_action_explicit}) associated with Brown-Henneaux modes (\ref{FG_diffeomorphism}) are responsible for the maximal Lypunov index at least in non-rotating BTZ. Note that since we have been cavalier about causality in our propagators \eqref{propagatorBTZ} we would not reproduce the $(t-|\phi|)$ behaviour in the exponent like \cite{Shenker:2014cwa}\footnote{If one does consider this then there will be possibly step functions multiplying each of the terms above.}.  
\subsection{Rotating BTZ}\label{rotating_btz}
Let's do the similar exercise for rotating BTZ where $\beta_\pm$ are complex parameters. Here in order to use the propagator in (\ref{propagator}) we would have to use a shifted Euclidean time in (\ref{tildecoord}) $\tilde{\tau}=(\tau+\mu \phi)/(1+\mu^2)$ for ordering the different Euclidean times. $i.e.$ for time ordered correlator we arrange $\tilde{\tau}_1>\tilde{\tau}_2>\tilde{\tau}_3>\tilde{\tau}_4$ while for out-of-time-ordered $\tilde{\tau}_1>\tilde{\tau}_3>\tilde{\tau}_2>\tilde{\tau}_4$. We would then fix the $\tilde{\tau}$ on a circle of period $\beta$. 
We will return later to the detail of how specifying the Euclidean time ordering in shifted Euclidean time does not effect the end result. 
\\\\
For the time-ordered case the exact Euclidean answer is given in the appendix (\ref{EuclideanTObtz}) and omitted here for the sake of brevity. As before we introduce Lorentzian time $\tilde{t}$  for the shifted coordinate $\tilde{\tau}$. We will infer the Lorenztian equivalent of (\ref{tildecoord}) later. We fix the Euclidean times to  $\tilde{\tau}_1=\beta,\tilde{\tau_2}=\beta/4,\tau_3=\beta/2,\tilde{\tau}_4=3\beta/4$ and further fix $\tilde{t}_1=\tilde{t}_2=\tilde{t}, \tilde{t}_3=\tilde{t}_4=0$ \& $\tilde{\phi}_1=\tilde{\phi}_2=0,\tilde{\phi}_3=\tilde{\phi}_4=\tilde{\phi}$. For fixed $\tilde{\phi}$ this corresponds to setting $\tau_1>\tau_2>\tau_3+\mu\tilde{\phi}>\tau_4+\mu\tilde{\phi} $. 
Having done this (\ref{EuclideanTObtz}) shows a polynomial growth in time.
\\\\
Similar expression for the {\it out-of-time-ordered } Euclidean answer is given in  (\ref{EuclideanOTObtz}).
Introducing Lorentzian times and fixing Euclidean times to $\tilde{\tau}_1=\beta,\tilde{\tau}_3=\beta/4,\tilde{\tau}_2=\beta/2,\tilde{\tau}_4=3\beta/4$ so as to compute OTOC; we then fix  $\tilde{t}_1=\tilde{t}_2=\tilde{t}, \tilde{t}_3=\tilde{t}_4=0$ \& $\tilde{\phi}_1=\tilde{\phi}_2=0,\tilde{\phi}_3=\tilde{\phi}_4=\tilde{\phi}$. This implies taking $\tau_1>\tau_3+\mu\tilde{\phi}>\tau_2>\tau_4+\mu\tilde{\phi}$. It is worth noting that this {\it out-of-time-ordering} is unaffected by the value of $\tilde{\phi}$.  Here we find the exponentially growing term in $\tilde{t}$ as
\be
\frac{\langle V_1(t,0) W_3(0,\phi) V_2(t,0) W_4(0,\phi) \rangle_{grav}}{\langle V_1(t,0) V_2(t,0) \rangle\langle W_3(0,\phi) W_4(0,\phi)\rangle}\sim  \frac{G_N\beta}{l}
\left[h_1h_2\cosh\left(\tfrac{2\pi(\tilde{t}-\tilde{\phi})}{\beta}\right)+\bar{h}_1\bar{h}_2\cosh\left(\tfrac{2\pi(\tilde{t}+\tilde{\phi})}{\beta}\right)\right]
\label{btzChaos}
\ee
Let's convert this back to $\{t,\phi\}$ by the Lorentzian version of (\ref{tildecoord}) $i.e.$
\begin{equation}
\tilde{x}^\pm=\tilde{t}\pm\tilde{\phi}=\frac{x^\pm}{(1\mp\mu_L)}\implies \phi=\tilde{\phi}-\mu_L\tilde{t},\,\,\,t=\tilde{t}-\mu_L\tilde{\phi}
\label{tildecoordLor}
\end{equation}
where we define the Lorenztian angular velocity as $\mu_L=i\mu=r_-/r_+$ as it would has risen in a Lorenztian bulk geometry, thus yielding\footnote{Here $\delta x^\pm=t\mp\phi$.}
\begin{equation}
\frac{G_N\beta}{l}
\left[h_1h_2\cosh\left(\tfrac{2\pi(t-\phi)}{\beta_+}\right)+\bar{h}_1\bar{h}_2\cosh\left(\tfrac{2\pi(t+\phi)}{\beta_-}\right)\right]
\end{equation}
Note, that the transformation (\ref{tildecoordLor}) is a conformal transformation of the boundary in the metric (\ref{Banados_metric}). The proper boundary coordinates along the lines of section 2 of \cite{Maldacena:2016upp} are $\{t,\phi\}$ which is what one must use to measure correlators in the boundary. It is clear from the last expression the Lypunov index for the each of the left and right moving modes  is governed by $\beta_\pm=\beta(1\mp\mu_L)$ in stead of $\beta$. Thus the Lypunov index for the 4pt OTOC would be $\lambda_L=2\pi/\beta_+>2\pi/\beta$ as it governs the fastest growth.
\\\\
A some what similar conclusion was reached in \cite{Stikonas:2018ane} with mutual information computed between the left and right intervals of $|\left.\text{TFD}\right\rangle$ corresponding to a rotating BTZ. This was computed both by computing mutual information $via$ R\'{e}nyi entropy in the 2D CFT with a chemical potential $\mu$ perturbed by a heavy operator, and from the bulk by employing the Ryu-Takayanagi prescription of minimal surfaces in a shock wave background. In \cite{Stikonas:2018ane} the symmetry between $\beta_\pm$ is broken by the spatial arrangement of the heavy operator relative to the entangling interval in question. This arrangement is such that only one of the modes with a smaller temperature effects the entangling region for positive times. For a different spatial arrangement one may seem to find that the scrambling time computed is governed by the higher of the 2 temperatures. 
\\\\
In the above computations we had chosen to order the Euclidean times in the shifted coordinate frame of $\tilde{\tau}=(\tau+\mu\phi)/(1+\mu^2)$ instead of ordering their $\tau$ coordinates. The reason why we still end up computing the desired correlator in the desired frame is as follows: The behaviour of the 4pt function is encapsulated in the sum conformal blocks. This in the limit of large central charge for holographic CFTs can be seen to be captured by only the Virasoro conformal block of the identity operator \cite{Roberts:2014ifa}. The conformal blocks are given by hypergeometric functions $F(a,b,c,\eta)$\footnote{Here $a,b,c$ do not encode bulk information and depend upon the conformal block in question.} that are known to have a branch point at $\eta=1$ where $\eta$ is the conformal cross ratio\footnote{Similar comments would hold true for the anti-holomorphic sector as the blocks factorize.}. It was shown in \cite{Roberts:2014ifa} that the behaviour of $\eta$ for large Lorentzian times depends on the ordering of the Euclidean times\footnote{Here $\eta$ is analytically continued to have both Lorentzian and Euclidean co-ordinates.}. For {\it out-of-time-ordering} of the Euclidean times, $\eta$ circles the branch point $\eta=1$ in the complex plane; while this is not the case when the Euclidean times are {\it time-ordered}. The authors of \cite{Roberts:2014ifa} then use the monodromy of $F(a,b,c,\eta)$ to compute it's behaviour for small $\eta$ thus obtaining the Lyapunov exponent for 2dim holographic CFTs. One can therefore use this behaviour of the cross ratio to discern if one is indeed computing {\it time-ordered} or {\it out-of-time-ordered} correlators.   
\\\\
We define the cross ratio $\eta$ as
\be
\eta=\frac{\omega_{12}\omega_{34}}{\omega_{13}\omega_{24}},\hspace{1cm}\omega=e^{-2\pi i z/\beta_-},\hspace{1cm}z=\tau	+ i\phi
\label{cross_ratio}
\ee
and similarly for the barred coordinates which are complex conjugates. $\eta$ can be analytically continued to include Lorentzian times by $\tau\rightarrow\tau -it$. The tilde coordinates are related to un-tilde coordinates by \eqref{tildecoord} $i.e.$
\be
\frac{z}{\beta_-}=\frac{\tilde{z}}{\beta},\hspace{0.2cm}\frac{\bar{z}}{\beta_+}=\frac{\tilde{\bar{z}}}{\beta}\,\,\,\implies\,\,\,\omega=e^{-2\pi i \tilde{z}/\beta},\hspace{0.2cm}\bar{\omega}=e^{2\pi i \tilde{\bar{z}}/\beta}
\label{tilde_coord_1}
\ee  
In all the above cases we choose $\tilde{\phi}_{1,2}=0,\tilde{t}_{3,4}=0$ \& $\tilde{t}_{1,2}=\tilde{t},\tilde{\phi}_{3,4}=\tilde{\phi}$, therefore for fixed values of Euclidean times $\tilde{\tau}_i$, $\eta(\tilde{t},\tilde{\phi})$ is a complex function. It was demonstrated in \cite{Roberts:2014ifa} that for $|\tilde{t}|>|\tilde{\phi}|>0$, the {\it out-of-time-ordered} cross-ratio $\eta_{oto}(\tilde{t},\tilde{\phi})$ circles the branch point $\eta=1$ as one varies $\tilde{t}$; while no such behaviour is found in the case for the {\it time-ordered} cross-ratio $\eta_{to}(\tilde{t},\tilde{\phi})$. In order to know how the shifted Euclidean time ordering effects the answer in the un-tilde frame we would like to see how $\eta$ (and $\bar{\eta}$) behaves as a function of $\{t,\phi\}$ for the respective orderings in $\tilde{\tau}_i$. 
\\\\
The cross-ratio $\eta$ for arbitrary $\tilde{\tau}_i$ looks like
\bea
\eta(\{\tilde{\tau}_i\},\tilde{t},\tilde{\phi})&=&\frac{(e^{-\frac{2\pi i}{\beta}\tilde{\tau}_{12}}-1)(e^{-\frac{2\pi i}{\beta}\tilde{\tau}_{43}}-1)}{(e^{-\frac{2\pi i}{\beta}\tilde{\tau}_{13}}e^{\frac{2\pi i}{\beta}(\tilde{t}+\tilde{\phi})}-1)(e^{-\frac{2\pi i}{\beta}\tilde{\tau}_{42}}e^{-\frac{2\pi}{\beta}(\tilde{t}+\tilde{\phi})}-1)}\cr
&=&\frac{(e^{-\frac{2\pi i}{\beta}\tilde{\tau}_{12}}-1)(e^{-\frac{2\pi i}{\beta}\tilde{\tau}_{43}}-1)}{(e^{-\frac{2\pi i}{\beta}\tilde{\tau}_{13}}e^{\frac{2\pi}{\beta_-}(t+\phi)}-1)(e^{-\frac{2\pi i}{\beta}\tilde{\tau}_{42}}e^{-\frac{2\pi}{\beta_-}(t+\phi)}-1)}.
\label{corssratio_framechange}
\eea
where in the second line above we have performed the Lorentzian coordinate transformation \eqref{tildecoordLor}. A similar expression holds for $\bar{\eta}$ with $\beta_-\rightarrow\beta_+$ and $t+\phi\rightarrow t-\phi$. Here $\beta\pm=\beta(1\mp\mu_L)$. We  plot the second expression in \eqref{corssratio_framechange} for different values of $\mu_L\in \{0,1\}$, $\tilde{\tau}_i=\beta-(i-1)\frac{3\beta}{4}+(i-1)\epsilon\frac{\beta}{4}$ and $\beta=2\pi$ with $\epsilon\in\{0,1\}$ for the {\it time-ordered} case. For the {\it out-of-time-ordered} case we exchange $\tilde{\tau}_2\leftrightarrow\tilde{\tau}_3$ as compared to the {\it time-ordered} case.  We find the same behaviour as seen in \cite{Roberts:2014ifa} for different values of chemical potential $\mu_L$ and $\epsilon$. Here $\epsilon$ varies the adjacent Euclidean time differences on the thermal circle linearly from $\beta/4$ to zero. Therefore even though we ordered shifted Euclidean times we end up computing the desired correlator in the un-tilde Lorentzian frame spanned by $\{t,\phi\}$\footnote{This might seem counter-intuitive at first,  but there is no precise way in which one can change frames when both Euclidean and Lorentzian space-time coordinates are present with non-zero chemical potential.}.
\\\\
One could further ask the question  whether a purely CFT computation of the form of \cite{Roberts:2014ifa} can recover(or corroborate) the results obtained for the rotating BTZ. This would require using the hypergeometric function $F(a,b,c,\eta)$  in the large central charge limit to incorporate not only temperature but also $\mu_L$ due to rotation. In \cite{Fitzpatrick:2015zha} it was shown that effect of introducing heavy operators at time-like infinities in 2dim CFT corresponds to evaluating the global conformal blocks on a new 2dim background  metric obtained by mapping the flat Euclidean space to a torus by\footnote{We are thankful to Justin David for having pointed this out.}
\be 
\omega=e^{-2\pi i z/\beta_-},\hspace{0.2cm}\bar{\omega}=e^{2\pi i \bar{z}/\beta_+}.
\label{flat_torus_euclid}
\ee 
where $\{\omega,\bar{\omega}\}$ are the Euclidean flat-space light-cone coordinates.
We leave this as an exercise for the near future.

\section{Results and Discussions}
The bulk understanding of how the Lyapunov index is $2\pi/\beta$ has as of yet always relied upon near the extremal property of an $AdS$ black hole exhibiting an $AdS_2$ throat. In some sense the back reaction of the scalars in the bulk gets the most contribution from this region. Any deviation in the $AdS_2$ geometry is captured in an action like the Jackiw-Teitielboim thus giving rise to a Schwarzian action.
What we have demonstrated here - at least in $AdS_3$; is that the Schwarzian arises even when one is far away from extremality. Moreover we find this as an effective action at the boundary of $AdS_3$ rather than at some screen in the interior.  It would be interesting to investigate how such an action can be arrived at for black holes in $AdS_{d>3}$, this would indeed give some understanding of the soft modes in higher dimensional large $N$ CFTs.
\\\\
In the probe approximation there is an inherent conformal symmetry in the bulk emanating from the asymptotic symmetries of $AdS_3$. This would correspond to the 2 copies of Virasoro algebra in the boundary CFT. Any arbitrary solution to the Einstein's equation with matter would  not have such a symmetry. Expanding perturbatively  about the probe approximation in orders of $G_N$ ($i.e.$ back reaction) breaks this symmetry spontaneously. The action (\ref{bulk_onshell_action_explicit}) therefore can be seen as the action cost associated with conformal transformations at the boundary of $AdS_3$  when one tries to go away from the probe approximation to linear order in $G_N$. 
\\\\
Extremality can be reached in the simplest possible manner by turning on charges for the $AdS$ black hole, the top down understanding of \citep{Maldacena:2016upp} in such a setting was explained in \citep{Nayak:2018qej} in $AdS_4$. Here the authors studied a probe uncharged scalar in the bulk thus having no dynamics for the gauge field. It would be interesting to analyse how the near horizon picture in \citep{Maldacena:2016upp} is reached for rotating geometries close to extremality in $AdS_{d>3}$. In \cite{Castro:2018ffi} 5d rotating Kerr geometries were analysed close to their near extremal limit in the near horizon throat region. There the authors have discovered a generalized JT action consisting of a dilaton and an additional scalar. 
\\\\
 In (\ref{btzChaos}) we take the view that $\beta_\pm$ are complex to begin with. The complex value of $\beta_\pm=\beta \mp i \Omega$ is required to make sense of the Euclidean BTZ metric as a real quantity. Further the $i\epsilon$\footnote{Here Euclidean time is $\tau$ instead of $\epsilon$.} prescription that we use requires us to first compute a Euclidean correlator and then analytically continue it to desired Lorentzian times. This is similar to the technique employed in \cite{Stikonas:2018ane} for computing the mutual entanglement from the 2D CFT, as computing the R\'{e}nyi entropy involves analytically continuing the Euclidean 4pt correlator $\langle \psi \sigma \tilde{\sigma} \psi^\dagger\rangle $\footnote{$\psi$ is the heavy operator generating the shock-wave in the BTZ while $\sigma$ is the twist operator. } to obtain the Lorentzian answer.
This requires making the left and right moving temperatures real: $\beta_\pm=\beta\mp\Omega$\footnote{In our analysis $\beta_\pm\sim\frac{1}{\sqrt{L_\pm}}\sim\frac{1}{\sqrt{M\pm J/l}}$ are associated with $x^\pm$ respectively.} when all Euclidean times have been put to zero. (\ref{btzChaos}) would therefore yield a growth in the scrambling time that would be greater than  the Lyapunov index of $2\pi/\beta$ due to the left moving (anti-holomorphic) modes for $x^+$ $i.e.$ $\frac{2\pi}{\beta-\Omega}$.
\\\\
The presence of rotation in the bulk implies a CFT with a chemical potential corresponding to angular momentum. 1d SYK and gauged-SYK models have been studied in the presence of a chemical potential \cite{Bhattacharya:2017vaz}. Here the Lypunov index computed was found to be bounded by $2\pi/\beta$. This also bodes well with the intuition that holding other charges fixed makes the system less chaotic. However the chemical potential present in such cases were associated to an internal symmetry and not a space-time symmetry.
\\\\
The analysis of section (4.2) for the case of rotating BTZ seems to yield a result in contradiction with the mathematical proof in \cite{Maldacena:2015waa}. The proof in section (4.1) of \cite{Maldacena:2015waa} is basically based on the maximal modulus theorem for a bounded holomorphic function. Here we try to give a simple understanding as to how one may try to reconcile the result of section (4.2) of this paper with that of \cite{Maldacena:2015waa}. 
\\\\
For the case of 2d CFTs one could argue a change in the Lyapunov index using the fact that the function of the form of $\frac{\langle V_1 V_2 W_3 W_4 \rangle}{\langle V_1 V_2 \rangle\langle W_3 W_4\rangle}$ of Virasoro primaries must be a scalar under conformal transformations. 
\be
\frac{\langle V_1 V_2 W_3 W_4 \rangle}{\langle V_1 V_2 \rangle\langle W_3 W_4\rangle}={\rm F}(x_{13},x_{24},x_{14}),\,\,\,{\rm where} \,\,\,x_{ij}=x_i-x_j
\ee
For the specific case of 2dim CFT dual to the Schwarzchild-$AdS_3$ this function was shown \cite{Shenker:2014cwa} to have a behaviour of 
\[
\frac{\langle V_1 V_2 W_3 W_4 \rangle_{oto}}{\langle V_1 V_2 \rangle\langle W_3 W_4\rangle}\sim 1 + \# G_Ne^{\frac{2\pi}{\beta}(t-|x|)}=
\begin{cases}
1 + \# G_Ne^{\frac{2\pi}{\beta}(t+x)} & {\rm for}\, x<0\\&\\
1 + \# G_Ne^{\frac{2\pi}{\beta}(t-x)} & {\rm for}\, x>0.
\end{cases}
\]
The finite diffeomorphism which takes Schwarzchild-$AdS_3$ to a generic rotating BTZ black- hole corresponds to a conformal transformation on the boundary of the form \eqref{tildecoordLor}
\be
\frac{ \delta x^\pm}{\beta}\rightarrow \frac{ \delta \tilde{x}^\pm}{\beta} =\frac{ \delta x^\pm}{\beta_\pm}
\label{tildecoordLor_2}.
\ee
where $\beta_\pm=(1\mp\mu_L)$ and $\delta x$ is the difference in any two of the coordinates in the correlator.
This would imply studying a holographic CFT with a chemical potential corresponding to rotation turned on.
The proof in section (4.1) of \cite{Maldacena:2015waa} relies crucially upon mapping the half strip of width $\beta/2$ in Euclidean time $\tau$ to a disk $via$ a conformal map \be w=\frac{1-\sinh\left[\tfrac{2\pi}{\beta}(t+i\tau)\right]}{1+\sinh\left[\tfrac{2\pi}{\beta}(t+i\tau)\right]}. \ee
This map has a periodicity under $\tau\rightarrow\tau+\beta$ which is exhibited by $\frac{\langle V_1 V_2 W_3 W_4 \rangle}{\langle V_1 V_2 \rangle\langle W_3 W_4\rangle}$. It was argued that this half strip is the minimum area of analyticity a function of the form $\frac{\langle V_1 V_2 W_3 W_4 \rangle}{\langle V_1 V_2 \rangle\langle W_3 W_4\rangle}$ can have. Smaller the width of the strip, greater would be the bound on chaos. However conformal transformation of the form \eqref{tildecoordLor_2} suggests that there exist 2dim holographic CFTs for which $\frac{\langle V_1 V_2 W_3 W_4 \rangle}{\langle V_1 V_2 \rangle\langle W_3 W_4\rangle}$ has a thinner region of analyticity of width $\beta_-/2=\beta(1-\mu_L)/2$ controlled by the chemical potential $\mu_L$. This suggests that repeating the analysis of \cite{Maldacena:2015waa} for systems at finite temperature and chemical potential might in certain cases yield a smaller region of analyticity in the Euclidean time direction.
\\\\
One could very well have guessed such a result simply by observing that the effective action (\ref{bulk_onshell_action_explicit}) doesn't mix the left and right movers, each of which have different inverse temperatures $i.e.$ $\beta$. The conformal blocks used in \cite{Roberts:2014ifa} to study chaos in 2dim holographic CFTs also exhibit such a holomorphic and anti-holomorphic factorization. From such a point of view it seems reasonable to expect that there would be 2 Lyapunov indices governed by the 2 temperatures; this point of view does not seem to be at odds with the statement of maximal chaos first proposed in \cite{Maldacena:2015waa}.
Therefore the maximal growth would be governed by the smaller of the two $i.e.$ $min[\beta_+,\beta-]$, while the surface gravity of the bulk would be related to the average  $\frac{\beta_++\beta_-}{2}$. 
However one can easily see that this would also lead to extremal BTZ metric exhibiting chaotic behaviour with the Lyapunov index being one of the 2 temperatures which is  non-vanishing. This would precisely be the temperature which would be greater than that given by the surface gravity for a generic rotating BTZ. 
This raises an important question in higher dimensional Kerr-$AdS$ black holes of the possibility of extremal chaos. 
It would be very interesting to see how these considerations would have to be modified when analysing rotating geometries in $AdS_{d>3}$ as unlike $AdS_3$ the bulk degrees of freedom of the metric would also participate in the dynamics.     
It is also worth noting that bulk explanations of the dynamics of chaotic behaviour using the 2dim Jackiw-Teitelboim model in the near horizon region of near extremal black holes \cite{Maldacena:2016upp} cannot be used to explain such extremal chaos. Therefore it would be interesting to find a similar JT like model(if possible) in the near horizon region of extremal and/or near extremal region to explain such a behaviour.  This would provide a useful tool in analysing such behaviour in higher dimensional Kerr-$AdS$ black holes.
\\\\
The result of section (4.2) is also validated by the analysis of mutual information for late times computed in the BTZ geometry subjected to a shock wave \cite{Stikonas:2018ane}. Here the author found the Lyapunov index to be related to the smaller of the two temperatures $i.e.$ $\lambda_L=\frac{2\pi}{\beta_-}$ in the conventions of this paper. The mutual information in the $|TFD\rangle$ state corresponding to an eternal BTZ subjected to a shock wave is computed in  \cite{Stikonas:2018ane} on the boundary by taking the limit of the R\'{e}nyi entropy, and in the bulk by employing the Ryu-Takayanagi prescription of minimal area. However the spatial arrangement of the heavy operator in \cite{Stikonas:2018ane} $w.r.t.$ the entangling region under consideration only sees the effect of one of the modes.       
\\\\
We have analysed the bulk on-shell action \eqref{bulk_onshell_action_explicit} upto quadratic order in fluctuations about a generic BTZ metric. It is in this approximation that the left and the right sectors seem to decouple and the scrambling times for the left and the right sectors are governed by their respective temperatures.
The full non-linear action \eqref{bulk_onshell_action_explicit} if expanded beyond quadratic order in fluctuations would indeed have terms which mix the left and the right moving sectors. It would be interesting to study how the three temperatures $\beta$ \& $\beta_\pm$ participate in  the process of reaching a thermal equilibrium at late times when the boundary CFT is excited. This would require a more detailed analysis for longer time periods than the scrambling times of the left and right sectors\footnote{The author is indebted to the referee for drawing attention to this. }. 
\\\\ 
The techniques used in this paper seem to be too well suited for $AdS_3$. As mentioned before that generalizing this to higher dimensional $AdS$ black holes would be interesting, it would nonetheless be easier to analyse the rotating BTZ along the lines of \cite{Shenker:2014cwa} by computing bulk eikonal scattering which seems to be an analysis suited for all dimensions\footnote{Barring the difficulty of computing bulk to boundary propagators for rotating black holes in $AdS_{d>3}$.}.   
\\\\
To conclude, this work also suggests that if the Lypunov index associated with rotating $AdS$ black holes in Einstein-Hilbert theory have maximal chaos, then for large $N$ thermal CFTs with chemical potential associated with angular momentum the chaos bound is $\lambda_L=\frac{2\pi}{\beta(1-\mu_L)}$. It would be interesting to find a more thorough generalization of the proof in \cite{Maldacena:2015waa} for large-$N$ CFTs with chemical potential associated with angular momentum.  
\section*{Acknowledgements}
The author is indebted to Gautam Mandal for fruitful discussions throughout the duration of this work. The author also benefited by earlier work with Gautam Mandal, Pranjal Nayak, Nemani Suryanarayana and Spenta Wadia. The author also acknowledges Adwait Gaikwad, Anurag Kaushal, R. Loganayagam, Pranjal Nayak, Ronak Soni, Shiraz Minwalla \& M. V. Vishal for their numerous discussions during the completion of this work. The author also acknowledges Abhijit Gadde and Justin David for giving inputs on this work. The author also found the academic environment of the Indian Strings Meet 2018 held in IISER Thiruvanathapuram conducive towards the completion of this work.
\section{Appendix}
The time ordered Euclidean answer for (\ref{BB}) corresponding to the rotating BTZ is
\bea
&&\frac{\langle V_1 V_2 W_3 W_4 \rangle_{grav}}{\langle V_1 V_2 \rangle\langle W_3 W_4\rangle}\Big|_{TO}=\cr
&&h_1h_2\left\lbrace\tfrac{8}{3}\left(\tfrac{\pi}{\beta(\mu-i)}\right)^6(z_{14}^4+z_{23}^4-z_{13}^4-z_{24}^4)\cot\left(\tfrac{\pi z_{12}}{\beta(1+i\mu)}\right)\cot\left(\tfrac{\pi z_{34}}{\beta(1+i\mu)}\right)\right.\cr
&&+\tfrac{16}{3}\left(\tfrac{i\pi}{\beta(\mu-i)}\right)^5\left[-3\pi z_{12}z_{34}(z_{13}+z_{24})\cot\left(\tfrac{\pi z_{12}}{\beta(1+i\mu)}\right)\cot\left(\tfrac{\pi z_{34}}{\beta(1+i\mu)}\right)\right.\cr
&&\left.\hspace{2.5cm}+(z_{13}^3+z_{23}^3-z_{14}^3-z_{24}^3)\cot\left(\tfrac{\pi z_{34}}{\beta(1+i\mu)}\right)+(-z_{13}^3+z_{23}^3-z_{14}^3+z_{24}^3)\cot\left(\tfrac{\pi z_{12}}{\beta(1+i\mu)}\right)\right]\cr
&&+\tfrac{8}{3}\left(\tfrac{\pi}{\beta(\mu-i)}\right)^4\left[-3(z_{13}^2+z_{14}^2+z_{23}^2+z_{24}^2)+2(3-\pi^2)z_{12}z_{34}\cot\left(\tfrac{\pi z_{12}}{\beta(1+i\mu)}\right)\cot\left(\tfrac{\pi z_{34}}{\beta(1+i\mu)}\right)\right.\cr
&&\hspace{2.5cm}\left.+3\pi(-z_{13}^2+z_{24}^2-z_{14}^2+z_{23}^2)\cot\left(\tfrac{\pi z_{12}}{\beta(1+i\mu)}\right)+3\pi(+z_{13}^2-z_{24}^2-z_{14}^2+z_{23}^2)\cot\left(\tfrac{\pi z_{34}}{\beta(1+i\mu)}\right)\right]\cr
&&+\tfrac{8}{3}\left(\tfrac{\pi}{\beta(\mu-i)}\right)^3\left[6i\pi (z_{13}+z_{24})+2i(\pi^2-3)\left(z_{34}\cot\left(\tfrac{\pi z_{34}}{\beta (1+i\mu)}\right)+z_{12}\cot\left(\tfrac{\pi z_{12}}{\beta (1+i\mu)}\right)\right)\right]\cr
&&\left.+\tfrac{16}{3}\left(\tfrac{\pi}{\beta(\mu-i)}\right)^2(\pi^2-3)\right\rbrace+ {\rm c.c.}\Big|_{h_1\rightarrow\bar{h}_1,h_2\rightarrow\bar{h}_2}
\label{EuclideanTObtz}
\eea
The above expression yields a polynomial expression in terms of the coordinates after analytically continuing to the Lorentzian times.
Similarly the out of time ordered Euclidean answer for the rotating BTZ case is
\bea
&&\frac{\langle V_1 V_2 W_3 W_4 \rangle_{grav}}{\langle V_1 V_2 \rangle\langle W_3 W_4\rangle}\Big|_{OTO}=\cr
&&h_1h_2\left\lbrace\tfrac{8}{3}\left(\tfrac{\pi}{\beta(\mu-i)}\right)^6(z_{14}^4+z_{23}^4-z_{13}^4-z_{24}^4)\cot\left(\tfrac{\pi z_{12}}{\beta(1+i\mu)}\right)\cot\left(\tfrac{\pi z_{34}}{\beta(1+i\mu)}\right)\right.\cr
&&+\tfrac{16}{3}\left(\tfrac{i\pi}{\beta(\mu-i)}\right)^5\left[\pi(z_{13}^3-z_{14}^3+z_{23}^3+z_{24}^3)\cot\left(\tfrac{\pi z_{12}}{\beta(1+i\mu)}\right)\cot\left(\tfrac{\pi z_{34}}{\beta(1+i\mu)}\right)\right.\cr
&&\left.\hspace{2.5cm}+(z_{13}^3+z_{23}^3-z_{14}^3-z_{24}^3)\cot\left(\tfrac{\pi z_{34}}{\beta(1+i\mu)}\right)+(-z_{13}^3+z_{23}^3-z_{14}^3+z_{24}^3)\cot\left(\tfrac{\pi z_{12}}{\beta(1+i\mu)}\right)\right]\cr
&&+\tfrac{8}{3}\left(\tfrac{\pi}{\beta(\mu-i)}\right)^4\left[-3(z_{13}^2+z_{14}^2+z_{23}^2+z_{24}^2)+2(3-\pi^2)z_{12}z_{34}\cot\left(\tfrac{\pi z_{12}}{\beta(1+i\mu)}\right)\cot\left(\tfrac{\pi z_{34}}{\beta(1+i\mu)}\right)\right.\cr
&&\hspace{2.5cm}\left.-3\pi(z_{13}^2-z_{24}^2+z_{14}^2+z_{23}^2)\cot\left(\tfrac{\pi z_{12}}{\beta(1+i\mu)}\right)-3\pi(-z_{13}^2+z_{24}^2+z_{14}^2+z_{23}^2)\cot\left(\tfrac{\pi z_{34}}{\beta(1+i\mu)}\right)\right]\cr
&&+\tfrac{8}{3}\left(\tfrac{\pi}{\beta(\mu-i)}\right)^3\left[6i\pi z_{14}+6i\pi z_{23}\cot\left(\tfrac{\pi z_{12}}{\beta (1+i\mu)}\right)\cot\left(\tfrac{\pi z_{34}}{\beta (1+i\mu)}\right)\right.\cr
&&\hspace{2.5cm}\left.+2(\pi^2-3)\left(iz_{34}\cot\left(\tfrac{\pi z_{34}}{\beta (1+i\mu)}\right)+iz_{12}\cot\left(\tfrac{\pi z_{12}}{\beta (1+i\mu)}\right)\right)\right]\cr
&&\left.-\tfrac{8}{3}\left(\tfrac{\pi}{\beta(\mu-i)}\right)^2\left[2(3-\pi^2)+3\pi\left(\cot\left(\tfrac{\pi z_{12}}{\beta(1+i\mu)}\right)+\cot\left(\tfrac{\pi z_{34}}{\beta(1+i\mu)}\right)\right)-3\pi\frac{\sin\left(\tfrac{\pi(z_{13}+z_{24})}{\beta(1+i\mu)}\right)}{\sin\left(\tfrac{\pi z_{12}}{\beta(1+i\mu)}\right)\sin\left(\tfrac{\pi z_{34}}{\beta(1+i\mu)}\right)}\right]\right\rbrace\cr
&&+{\rm c.c.}\Big|_{h_1\rightarrow\bar{h}_1,h_2\rightarrow\bar{h}_2}
\label{EuclideanOTObtz}
\eea 
Introducing Lorentzian times and fixing $\tau_1=\tilde{\beta}-\mu \phi_1,\tau_3=\tilde{\beta}/4-\mu\phi_3,\tau_2=\tilde{\beta}/2-\mu\phi_2,\tau_4=3\tilde{\beta}/4-\mu\phi_4$ and further fixing $t_1=t_2=t, t_3=t_4=0$ \& $\phi_1=\phi_2=0,\phi_3=\phi_4=\phi$. Here we find the exponentially growing term in $t$ as
\be
\frac{\langle V_1(t,0) W_3(0,\phi) V_2(t,0) W_4(0,\phi) \rangle_{grav}}{\langle V_1(t,0) V_2(t,0) \rangle\langle W_3(0,\phi) W_4(0,\phi)\rangle}\sim  \frac{G_N\beta}{l}
\left[h_1h_2\cosh\left(\tfrac{2\pi(t-\phi)}{\beta_+}\right)+\bar{h}_1\bar{h}_2\cosh\left(\tfrac{2\pi(t+\phi)}{\beta_-}\right)\right]
\ee

\bibliographystyle{JHEP.bst}
\bibliography{bulk_syk_soft_modes.bib}
\end{document}